\documentclass[twocolumn,fleqn,usenatbib]{aastex63}
\usepackage{mathptmx}
\usepackage[T1]{fontenc}
\usepackage{ae,aecompl}

\usepackage{graphicx}   
\usepackage{amsmath}    
\usepackage{amssymb}    

\submitjournal{ApJ}

\shorttitle{CVs in ASAS-SN}
\shortauthors{Kawash et al.}

\begin{document}

\title[ASAS-SN Population Study]{Classical Novae Masquerading as Dwarf Novae?\\
Outburst Properties of Cataclysmic Variables with ASAS-SN}

\correspondingauthor{Adam Kawash}
\email{kawashad@msu.edu}

\author[0000-0002-0786-7307]{A.\ Kawash}
\affiliation{Center for Data Intensive and Time Domain Astronomy, Department of Physics and Astronomy, Michigan State University, East Lansing, MI 48824, USA}

\author{L.\ Chomiuk}
\affiliation{Center for Data Intensive and Time Domain Astronomy, Department of Physics and Astronomy, Michigan State University, East Lansing, MI 48824, USA}

\author{J.\ Strader}
\affiliation{Center for Data Intensive and Time Domain Astronomy, Department of Physics and Astronomy, Michigan State University, East Lansing, MI 48824, USA}

\author{E.\ Aydi}
\affiliation{Center for Data Intensive and Time Domain Astronomy, Department of Physics and Astronomy, Michigan State University, East Lansing, MI 48824, USA}

\author{K.~V.\ Sokolovsky}
\affiliation{Center for Data Intensive and Time Domain Astronomy, Department of Physics and Astronomy, Michigan State University, East Lansing, MI 48824, USA}
\affiliation{Sternberg Astronomical Institute, Moscow State University, Universitetskii~pr.~13, 119992~Moscow, Russia}

\author{T.\ Jayasinghe}
\affiliation{Department of Astronomy, The Ohio State University, 140 West 18th Avenue, Columbus, OH 43210, USA}

\author{C. S. \ Kochanek}
\affiliation{Department of Astronomy, The Ohio State University, 140 West 18th Avenue, Columbus, OH 43210, USA}
\affiliation{Center for Cosmology and Astroparticle Physics, The Ohio State University, 191 W. Woodruff Avenue, Columbus, OH 43210, USA}

\author{P.\ Schmeer}
\affiliation{Bischmisheim, Am Probstbaum 10, 66132 Saarbrücken, Germany}

\author{K.\ Z.\ Stanek}
\affiliation{Department of Astronomy, The Ohio State University, 140 West 18th Avenue, Columbus, OH 43210, USA}
\affiliation{Center for Cosmology and Astroparticle Physics, The Ohio State University, 191 W. Woodruff Avenue, Columbus, OH 43210, USA}

\author{K.\ Mukai}
\affiliation{CRESST II and X-ray Astrophysics Laboratory, NASA/GSFC, Greenbelt, MD 20771, USA}
\affiliation{Department of Physics, University of Maryland, Baltimore County, 1000 Hilltop Circle, Baltimore, MD 21250, USA}

\author{B.\ Shappee}
\affiliation{Institute for Astronomy, University of Hawai`i at M\=anoa, 2680 Woodlawn Dr., Honolulu 96822, USA}

\author{Z.\ Way}
\affiliation{Department of Astronomy, The Ohio State University, 140 West 18th Avenue, Columbus, OH 43210, USA}

\author{C.\ Basinger}
\affiliation{Department of Astronomy, The Ohio State University, 140 West 18th Avenue, Columbus, OH 43210, USA}

\author{ T.~W.-S.\ Holoien}
\affiliation{Carnegie Observatories, 813 Santa Barbara Street, Pasadena, CA 91101, USA}

\author{J.~L.\ Prieto}
\affiliation{N\'ucleo de Astronom\'ia de la Facultad de Ingenier\'ia y Ciencias, Universidad Diego Portales, Av. Ej\'ercito 441, Santiago, Chile}
\affiliation{Millennium Institute of Astrophysics, Santiago, Chile }

\begin{abstract}

The unprecedented sky coverage and observing cadence of the All-Sky Automated Survey for SuperNovae (ASAS-SN) has resulted in the discovery and continued monitoring of a large sample of Galactic transients. The vast majority of these are accretion-powered dwarf nova outbursts in cataclysmic variable systems, but a small subset are thermonuclear-powered classical novae. Despite improved monitoring of the Galaxy for novae from ASAS-SN and other surveys, the observed Galactic nova rate is still lower than predictions. One way classical novae could be missed is if they are confused with the much larger population of dwarf novae. Here, we examine the properties of 1617 dwarf nova outbursts detected by ASAS-SN and compare them to classical novae. We find that the mean classical nova brightens by $\sim$11 magnitudes during outburst, while the mean dwarf nova brightens by only $\sim $5 magnitudes, with the outburst amplitude distributions overlapping by roughly 15$\%$. For the first time, we show that the amplitude of an outburst and the time it takes to decline by two magnitudes from maximum are positively correlated for dwarf nova outbursts. For classical novae, we find that these quantities are negatively correlated, but only weakly, compared to the strong anti-correlation of these quantities found in some previous work. We show that, even if located at large distances, only a small number of putative dwarf novae could be mis-classified classical novae suggesting that there is minimal confusion between these populations. Future spectroscopic follow-up of these candidates can show whether any are indeed  classical novae.

\end{abstract}

\keywords{Classical novae (251), Dwarf novae (418), Novae (1127), Cataclysmic variable stars (203), White dwarf stars (1799)}

\section{Introduction} \label{sec:intro}

Interacting binary systems that consist of a white dwarf accreting material from a close companion star 
are known as Cataclysmic Variables (CVs). The secondary, usually a low-mass main-sequence star, transfers matter 
through Roche-lobe overflow, which forms an accretion disk around the white dwarf (see \citealt{warner_1995} and \citealt{hellier_2001} 
for reviews). 

A Dwarf Nova (DN) outburst is a common event that occurs in a CV, and
is generally thought to be caused by a thermal instability in the accretion disk (see \citealt{Hameury_2020} for a review). 
The disk rapidly transitions from neutral to ionized, leading to a sudden increase in disk viscosity and
mass-accretion rate, and a dramatic brightening of the accretion disk \citep{hellier_2001}.
%
DNe are one of the most common types of 
Galactic transients, with new objects being discovered generally every week (see \citealt{kiw20} 
and previous papers for examples). Individual systems will typically outburst every 20--300 days \citep{osaki01}. 
The peak absolute magnitude of a DN depends mostly on the physical size and inclination of the accretion disk, and ranges from M$_{V\rm{,max}}$ $\approx$ 7 to 2 mag,
with long orbital period systems viewed face-on producing
brighter outbursts \citep{2004AJ....127..460H, patterson11}. 

A Classical Nova (CN) is another type of event that occurs in a CV, and is caused by a thermonuclear runaway on the surface of the white dwarf (see \citealt{be08} for a review). Accreted material builds up on the surface of the white dwarf over time, until a critical pressure is reached, which triggers explosive thermonuclear burning and the puffing up and expulsion of the accreted envelope. 
Recent studies of CNe in M31 with well-constrained luminosities show that the absolute magnitude at peak brightness can range from M$_V \approx -4$ to $-10$ mag, much more luminous than DNe \citep{shafter_2017}. This is consistent with early estimates of the Galactic nova luminosity function \citep{mclaughlin45} even with more precise distances from \textit{Gaia} DR2.\cite{di20}.
Outbursts of CNe are expected to recur on a timescale that depends on the white dwarf mass and accretion rate \citep{Yaron_2005}, and if a CN has been observed to erupt more than once, it is referred to as a recurrent nova. 
Of the ten recurrent novae known in our Galaxy, the recurrence time ranges from 10 to 80 years \citep{Schaefer_2010}. 
However, in other galaxies, more rapidly recurring novae are being discovered \citep{Darnley_2019}, with a nova in M31 that has been found to recur every year \citep{dhh16}, and nova LMC~1968 recently exhibiting a four year recurrence time \citep{Kuin_2020, Page_2020}. 
For the purposes of this work,  we are interested in both recurrent and singular classical novae and do not distinguish between the two classes, considering them both thermonuclear-powered CNe.

There have been many estimates of the Galactic nova rate (see \citealt{di20} for a review), but it remains poorly constrained. Recently, \citep{shafter_2017} derived a rate of $50_{-23}^{+31}$ novae per year from Galactic observations and a rate between $\sim50$ and $\sim70$ novae per year from extragalactic observations. These rates, though mutually consistent, are larger than previous estimates, and significantly higher than the discovered rate.
Historically, amateur astronomers have played a leading role in the discovery and observations of CN outbursts, and found only a small fraction of the predicted population. 
The average number of discovered Galactic novae increased from about 3 per year in the mid 20th century (when many discoveries were made visually) to 4 per year in the 1980s and 1990s (when film photography was often used) to 8 per year in the 2000s and 2010s (when digital cameras became widely available)\footnote{Up-to-date lists of novae may be found at \url{https://asd.gsfc.nasa.gov/Koji.Mukai/novae/novae.html} and \url{https://github.com/Bill-Gray/galnovae}}. Amateur observers use a  variety of equipment, which often include an 
astronomical CCD camera attached to a telephoto lens or small telescope with typical detection limits down to $V\approx 12$ mag.
As amateur observations do not systematically cover the entire sky down to a well defined limiting magnitude, 
one explanation for the discrepancy between the number 
of predicted and discovered CNe in the Galaxy is that most CNe eruptions go undiscovered. 
To test this possibility, a deep wide-field survey with high observing cadence is needed. Fortunately, such a survey now exists.


The All-Sky Automated Survey for SuperNovae (ASAS-SN) is the only survey to date observing the entire night sky 
with nearly nightly cadence \citep{spg14,kss17}.
Early ASAS-SN observations were conducted at two facilities in Hawaii and Chile, using a \textit{V} filter with a few day cadence down to a depth of \textit{V} $\approx$ 17\,mag. In 2017, ASAS-SN added facilities in Texas, South Africa, and an additional facility in Chile, switched to observing in a \textit{g} filter down to a median depth of \textit{g} $\approx$ 18.5\,mag and became able to observe the entire night sky (including the Galactic plane) with nearly nightly cadence \citep{kss17,jsk19oct}.
The primary goal of ASAS-SN is to discover bright, extragalactic supernovae,
but due to the all-sky nature of the survey, there are a wide variety of transients discovered, including CV outbursts. 
The observing capabilities of ASAS-SN make it uniquely suited for monitoring fast outbursts from CVs brighter than \textit{g}~$\approx$ 18\,mag, considerably deeper than most amateur observations. Various models from \cite{shafter_2017} predict anywhere from 30 to 110 CNe brighter than \textit{V} $\approx$ 18 in the Galaxy each year, but so far ASAS-SN observations have yielded no large increase in the number of discovered novae.

With no increase in the discovery rate. another explanation must exist if the rate estimates are correct. One possibility is that CNe are being confused with the more numerous DN outbursts. Due to the high frequency of nearby DN outbursts, it is not feasible to obtain a classification spectrum of even a substantial fraction of DN candidate outbursts in the Galaxy. This problem will only be exacerbated by next generation time domain surveys like the Large Synoptic Survey Telescope (LSST; \citealt{lsst19}). The default assumption for a fainter CV outburst (\textit{g}$_{\rm peak} >$ 13\,mag) is that it is a DN; this is usually a safe assumption since a Galactic CN should be very bright, even when observed on the other side of the Galaxy unless the  dust extinction is very high.

The main goal of this work is to investigate the possibility that some Galactic CNe are being mistaken for DNe by inspecting the outburst properties of a large sample of both types of transients. 
For extragalactic CNe, the association with a well-studied nearby galaxy means that the CN distances--- and therefore absolute magnitudes--- are well constrained. Although the luminosity function of CNe is reasonably well-measured \citep[e.g.,][]{shafter_2017}, it has limited utility in the Galaxy where nova distances are usually poorly constrained; many Galactic CN progenitors are too faint in quiescence or too distant
for an accurate parallax measurement even with \textit{Gaia} \citep{Shaefer18,sg19}.
However, photometry of the field prior to outburst often exists for Galactic events, 
making it possible to estimate how much the object brightened during outburst---typically called the outburst amplitude. The outburst amplitude has the potential to be a powerful discriminant between CNe and other transients but is relatively little studied. 

Significant effort that has been invested in understanding the potential relationship between absolute magnitude of peak outburst and its rate of decline for CNe 
(MMRD; \citealt{Capaccioli+89, DellaValle&Livio95, kck11,sdl17}). 
There are far fewer studies of the relationship between outburst amplitude and decline time for CNe. 
\cite{warner87} found that CNe exhibit outburst amplitudes ranging from 8 to 15 magnitudes in \textit{V}-band, 
with large-amplitude CNe fading quickly and small-amplitude CNe sometimes taking years to fade back to quiescence, and 
this relationship has been used to identify potential recurrent nova candidates from the sample of known CNe
\citep{ps14}. Recurrent novae are expected to occur on massive white dwarfs and have short decline times \citep{Yaron_2005}. 
Given the large luminosity differences between CNe and DNe, this relationship could also be useful in identifying potential CNe 
candidates hiding in the large population of DNe. DNe typically have outbursts that last roughly a week with amplitudes of 2--5\,mag, 
much lower than CNe. 
However, WZ~Sge type DNe \citep{orr80,kato15,hellier_2001,1995ApJ...439..337H} show rare (once in decades)
accretion-powered superoutbursts with amplitudes reaching 9 magnitudes and lasting for weeks \citep[e.g.,][]{2020PASJ...72...49T}. 
Although the vast majority of DNe should have lower outburst amplitudes than CNe, WZ~Sge type superoutbursts could be confused with CNe if only the outburst amplitude but not the absolute magnitude is known. It is a goal of this paper to better understand this potential for confusion, and to investigate possibilities for alleviating it in order to more confidently identify CNe.

Recently, time-domain surveys have found large numbers of DN outbursts due to their high frequency. A few examples of these include the Sloan Digitial Sky survey (SDSS: \citealt{gds09}), the Catalina Real-time Transient Survey (CRTS: \citealt{ckk16}), and the Optical Gravitational Lensing Experiment (OGLE; \citealt{mup15}). These have resulted in large sample studies of a multitude of DN outburst properties, but we have found no previous discussion of the relationship between outburst amplitude and decline time from maximum for DNe.

In this work, we focus on the relationship between outburst amplitude and decline time as a potential tool for distinguishing CNe from DNe. To do this, we estimate the outburst properties of DNe, along with CNe that have erupted since ASAS-SN started observing in 2013 and compare the two samples. 
In Section \ref{sec:methods}, we describe how the sample of CVs was obtained, how the light curves were generated, and how the various outburst properties were measured.
In Section~\ref{sec:results}, we present the outburst properties of the CN and DN populations, 
fit the distributions of outburst amplitudes and decline times, measure the correlation between these two properties, 
and discuss the observable differences between the two types of outbursts. We then assess in Section~\ref{sec:hiding} whether CNe could be hiding amongst DNe, and how we can ensure in the future that the two types of transients are not confused.

\section{Methods}
\label{sec:methods}

\subsection{Catalog}
\label{sec:cat}
The list of CVs analyzed in this work was obtained from the AAVSO International Variable Star Index (VSX; \citealt{whp06}), which contains the most up-to-date and comprehensive list of  known CVs, including CVs discovered by ASAS-SN. VSX was queried using \texttt{TAPVizieR} \citep{los13} for any objects 
flagged as\footnote[2]{\url{https://www.aavso.org/vsx/index.php?view=about.vartypes}}:
\begin{itemize}
\item U Geminorum-type variables (``UG" flag), including all the sub-classes in the VSX catalog. These are CVs that have been typed as DNe.
\item DQ Herculis-type variables (``DQ" flag), which are CVs with intermediate-strength magnetic fields, and are also known as intermediate polars. Given the right orbital period, accretion rate, and magnetic field strength, these systems can still produce DN outbursts \citep{hl17}.
\item CVs of unknown type (``CV" flag). These are often CVs that have recently been discovered in surveys like ASAS-SN, and which have not yet been assigned a type in VSX.
\end{itemize}
A total of 9333 objects had these flags in the VSX catalog at the time the catalog was queried (December 2019).
There were a total 62 CN outbursts discovered in the Galaxy between January 2013 and April 2020. The positions of these CNe, like the sample of DNe, were obtained from VSX.

\subsection{Light curves}
\label{sec:lc}
\begin{figure}
\begin{center}
 \includegraphics[width=0.48\textwidth]{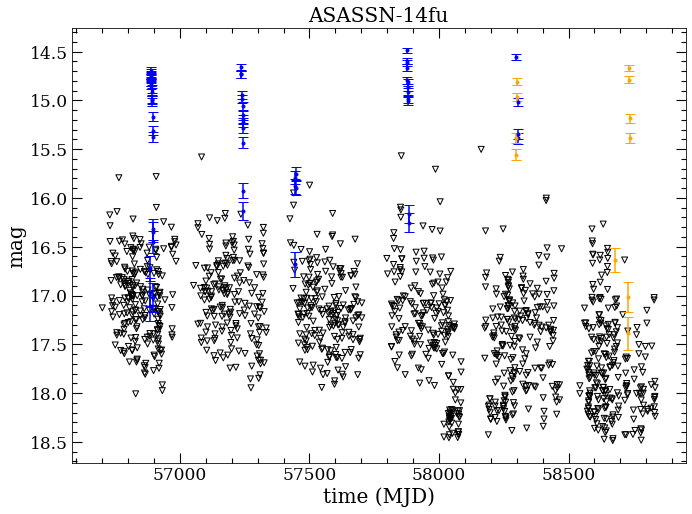}
\caption{The ASAS-SN light curve of the dwarf nova ASASN-14fu. The $\geq$5-sigma detections are shown for \textit{V}-band and \textit{g}-band observations in blue and orange, respectively. The black triangles denote 5$\sigma$ upper limits derived from non-detections, and the gaps in the data are due to seasonal the Solar constraints.}
\label{Fig:lc_ex}
\end{center}
\end{figure}

Image-subtraction light curves were generated using ASAS-SN observations for all objects in our sample following the procedures described in \citet[][see also \citealt{alard98,alard00}]{jks18,jt19}. ASAS-SN light curves for most fields outside of the Galactic plane span back to 2013. In 2017, ASAS-SN switched from observing in a \textit{V}-band filter to a \textit{g}-band filter and started more regularly monitoring the Galactic plane. For the purposes of our analysis, \textit{g}-band is used as the standard filter; the conversions of \textit{V}-band measurements to \textit{g}-band are outlined in Section~\ref{sec:filter}. An example of an ASAS-SN light curve for a DN is shown in Figure \ref{Fig:lc_ex} and additional light curves are shown in Figure \ref{Fig:nova_cand_lc}, Figure \ref{Fig:nova_cand_lc2}, and Figure \ref{Fig:nova_lcs}. 

Image-subtraction photometry is preferred over aperture photometry for studying CV outbursts. Reference images are created using the best images of each field with outlier rejection before the final average, which automatically rejects any outbursts. By subtracting this reference image from individual epochs, any flux at the position of a CV in the individual observations should be from an outburst. Contaminating flux from nearby stars, a problem given ASAS-SN's angular resolution (2 pixel full width at half maximum = 16 arcseconds), is removed by the subtraction, although bright stars are not always subtracted cleanly.  All the light curves used to measure the outburst properties were inspected for contamination. Some CVs within a few pixels of a bright star (\textit{g} $<$ 14\,mag) show artifacts due to reference image subtraction errors. These were flagged and ultimately dropped, leading to the elimination of $\sim$ 2\% of the sample. Higher resolution photometry of the environment was provided by the Panoramic Survey Telescope and Rapid Response System 
(Pan-STARRS; \citealt{cmm16}). 

ASAS-SN light curves for objects near the edge of a detector chip can have problems. These sources often lie in the overlap regions between fields, so data from one camera were flagged if the median magnitude was more than two magnitudes different from other cameras or if the flux limit of the ASAS-SN image was much lower than expected based on the observation duration.  

Light curves of Galactic CNe that have erupted since 2013 were also generated using image-subtraction photometry from ASAS-SN. These data were combined with \textit{V}-band observations from the American Association of Variable Star Observers (AAVSO; \citealt{Kafka_2020}) to increase the cadence and expand the sensitivity of our analysis for the CNe brighter than the saturation limit of
ASAS-SN (\textit{g} $\approx$ 10\,mag). The AAVSO data were visually inspected, and observations from individual observers were discarded if they were inconsistent with data from other contributors. These erroneous observations, though rare, likely occur when one object is mistaken for another in a crowded field.

\subsection{Outburst Peak Magnitude and Decline Time}
\label{sec:peakmag_t2}
Various aspects of the outburst can be measured directly from the light curve. To measure the maximum brightness, it is common to smooth the light curves of CNe \citep[e.g.,][]{bh08}. This allows jitters and short flares to be ignored when estimating the peak. However, for the purposes of this work, we define the peak brightness simply as the brightest observation in the light curve, as done by \cite{ssh10}. For most objects, the cadence of ASAS-SN provides observations very close to maximum brightness, 
but for transients evolving on a timescale less than a day, 
the maximum brightness can be underestimated. Also, outbursts that are discovered immediately after a field emerges from its Solar conjunction can have significantly underestimated peak brightness.

Another quantity that we are able to measure directly from the light curve is the decline time, $t_2$, defined as the time in days it takes for the light curve to decline by two magnitudes from maximum brightness. For DN outbursts, this is relatively straightforward, as they typically exhibit smooth declines, though we consider any plateaus in the light curve after maximum brightness to be part of the decline. CN light curves can exhibit jitters, flares, and cusps (see Figure \ref{Fig:nova_lcs} and \citealt{ssh10}), which can cause $t_2$ to change depending on the definition (e.g., first decline by two magnitudes versus final fade by two magnitudes). We define $t_2$ as the last time the light curve drops below two magnitudes from maximum in order to be consistent with the estimates
by \cite{ssh10}. 

To measure $t_2$, we first assumed that the brightest detection was the peak of an outburst. Then, we required that the decline have at least two detections separated by more than one hour to automatically eliminate satellite trails and asteroids. Next, we required that all data used to measure $t_2$ be brighter than the independently measured quiescent magnitude (discussed in \S\ref{sec:quimag}). This eliminates objects with outburst amplitudes less than 2 magnitudes, but was necessary to distinguish outbursts, CV variability in quiescence, and contamination from nearby bright stars. We also required that there is no gap between consecutive observations longer than 40 days to eliminate artificially extended $t_2$ values due to Solar conjunctions. Linear interpolation between the two data points above and below the two-magnitude threshold was used to estimate $t_2$.

We are able to tightly constrain $t_2$ when ASAS-SN observations are able to detect the outburst below the two-magnitude threshold. In these cases, we consider this a measurement of $t_2$. For some faint and fast outbursts close to the ASAS-SN detection limit, the outburst decline is not tracked all the way to the two magnitude threshold, but a subsequent non-detection places a limit fainter than the two-magnitude threshold. In this case, we limit t$_2$ to be bounded by these two epochs.

\subsection{Outburst Amplitude}\label{sec:amp}
Observations of the brightest outburst of a DN from ASAS-SN were combined with observations from The Pan-STARRS $3\pi$ Steradian Survey \citep{cmm16} of the same object in quiescence to estimate the amplitude of outburst. We estimate the quiescent magnitude of the CNe in the same way for those in the observing field of Pan-STARRS (declination $> -30^{\circ}$); otherwise we use \textit{Gaia} DR2 photometry \citep{bvp18}  for CNe that erupted after \emph{Gaia} DR2 observations were completed (2016 May 23). The details of the quiescent magnitude measurements are discussed in \S\ref{sec:quimag} and \S\ref{sec:catmatch}.

If an object is unambiguously detected in quiescence, we make a measurement of its outburst amplitude. However, if an object is clearly not detected (no match within four arcseconds for DNe and 2 arcseconds for CNe), we place a lower limit on its outburst amplitude. The outburst amplitude we estimate is simply the difference between the peak magnitude of the outburst detected by ASAS-SN or AAVSO observations and the magnitude obtained from the Pan-STARRS or Gaia photometry catalogs, after correcting for filter transformations (\S \ref{sec:filter}). If a CN was detected immediately after Solar conjunction, it is likely that the peak brightness was missed (See Figure \ref{Fig:nova_lcs}). We expect this is only an issue for CNe, since they can still be detected in outburst months after eruption. For these CNe, we place a lower limit on the outburst amplitude and an upper limit on $t_2$.  

\begin{figure}
\begin{center}
 \includegraphics[width=0.48\textwidth]{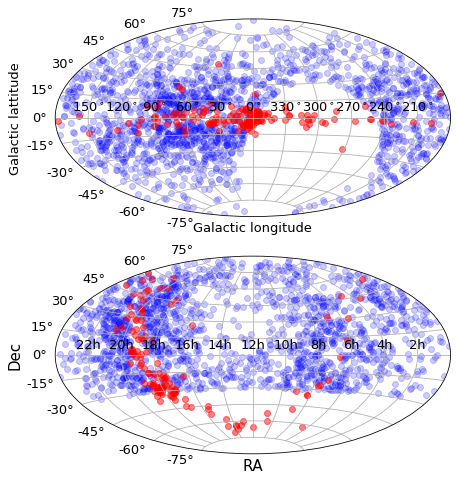}
\caption{Galactic (top) and equatorial  (bottom) coordinate positions of CVs with outburst property estimates. Dwarf novae are shown in blue, and classical novae are shown in red. The gap in the data for dwarf novae is due to the survey limits of Pan-STARRS}
\label{Fig:map_comb}
\end{center}
\end{figure}

\section{Results}
\label{sec:results}

\subsection{Detected DN and CN Outbursts} 
In total, we find 2688 DNe with outbursts that declined by at least two magnitudes from maximum in the ASAS-SN data, around 30$\%$ of all DNe in VSX. This does not test the discovery and classification of DNe in ASAS-SN, as we have only searched for outbursts from known DNe discovered by a variety of surveys and techniques. In order to be detected in our analysis, a DN needs to have gone into outburst in a field ASAS-SN regularly monitored (the entire sky since 2017), and to have reached a peak outburst magnitude in the range \textit{g} $\approx$ 10--16\,mag (a more detailed analysis of our detection capabilities is discussed in Section~\ref{sec:selection}). From the subset of objects with detected outbursts, 1791 objects have declination greater than $-30^{\circ}$, and are therefore in Pan-STARRS. We are able to unambiguously estimate or place a limit on the quiescent brightness for 1617 of these. The measured properties of these DNe are presented in Table \ref{table:dn_results} (the entirety of which is available online in a machine-readable format).

\begin{deluxetable*}{lcccccrcrr}
\tabletypesize{\small}
\tablecolumns{10}
\tablewidth{0pt}
\tablecaption{ Outburst Properties of Dwarf Novae \label{table:dn_results}}
\tablehead{
\colhead{Name} & 
\colhead{RAJ2000} & 
\colhead{DEJ2000} & 
\colhead{Peak} &
\colhead{Amp.} & 
\colhead{Amp. Flag} & 
\colhead{t$_2$} & 
\colhead{t$_2$ Flag} &
\colhead{t$_{2,low}$} & 
\colhead{t$_{2,up}$}\\
& \colhead{hms} & 
\colhead{dms} & 
\colhead{mag} &
\colhead{mag} & 
\colhead{boolean} & 
\colhead{days} & 
\colhead{boolean} &
\colhead{days} & 
\colhead{days} }
\startdata
ASASSN-18xt & 2:25:06.37 & 8:06:38.6 & 14.1 & 6.4 & 1 & 12.1 & 0 & 10.9 & 16.2 \\
CSS 091106 023638+111157 & 2:36:37.98 & 11:11:56.5 & 15.1 & 4.9 & 1 & 12.9 & 0 & 9.0 & 16.1 \\
TCP J03005508+1802290 & 3:00:55.05 & 18:02:28.7 & 12.1 & 3.4 & 1 & 11.9 & 1 & 11.4 & 11.9 \\
MLS 130110 034256+171739 & 3:42:56.18 & 17:17:40.1 & 16.0 & 4.7 & 1 & 3.8 & 0 & 3.7 & 5.1 \\
MLS 130302 035906+175034 & 3:59:05.90 & 17:50:34.5 & 16.3 & 2.4 & 1 & 18.4 & 0 & 6.0 & 21.0 \\
CSS 081118 041139+232220 & 4:11:38.58 & 23:22:20.3 & 15.0 & 4.6 & 1 & 11.0 & 0 & 7.0 & 13.9 \\
CSS 081107 033104+172540 & 3:31:04.44 & 17:25:40.2 & 15.5 & 4.9 & 1 & 5.9 & 0 & 4.0 & 7.0 \\
CSS 081107 033556+191119 & 3:35:55.78 & 19:11:19.1 & 15.6 & 5.3 & 1 & 14.2 & 0 & 6.9 & 16.1 \\
CSS 090213 033031+201402 & 3:30:31.41 & 20:14:01.2 & 15.6 & 4.3 & 1 & 5.0 & 0 & 4.5 & 5.0 \\
V0701 Tau & 3:44:01.97 & 21:57:07.4 & 15.2 & 6.4 & 1 & 16.3 & 0 & 15.0 & 19.9
\enddata
\tablecomments{Names, positions, peak apparent brightness, amplitude of outburst, and t$_2$ for the dwarf novae in our sample. The Amp. Flag columns equals 1 when we are able to make a measurement of the outburst amplitude and 0 when we are able to place a lower limit. The t$_2$ Flag column is 1 when were are able to detect the object below the two magnitude threshold and 0 when there is only a non-detection below this threshold. When t$_2$ Flag = 0, the value listed for t$_2$ is likely larger than the true value. The t$_{2,low}$ column gives the time until last detection above the two magnitude threshold and the t$_{2,up}$ column gives the time until the first detection or non-detection below this threshold.  These last two columns are lower and upper limits on t$_2$, respectively. The first 10 dwarf novae are shown here and the entirety of the this table is available in a machine readable format in the electronic paper.}
\end{deluxetable*}

By combining data from ASAS-SN and AAVSO, we are able to measure or place a limit on $t_2$ for 50 CNe. We are able to unambiguously estimate the quiescent brightness for 40 of these objects. The measured properties of these CNe are presented in Table \ref{table:cn_results} in Section \ref{sec:nova_results}. In order to make a more robust comparison between DN and CN outbursts, previous CN outburst estimates and limits were also obtained from \cite{ssh10}.  This yielded an additional 92 CNe for the sample,
bringing the total number of CN outbursts studied to 132.

The positions of both the DNe and CNe are shown in Galactic and equatorial coordinates in Figure \ref{Fig:map_comb}. The DNe in our sample are restricted to the Pan-STARRS observing field, but we also used \emph{Gaia} to have full sky coverage for the CNe. The CNe are generally restricted to within several degrees of the Galactic plane,
as expected if CVs track the stellar mass density of the Galaxy 
\citep[e.g.,][]{shafter_2017}. However, as DNe are likely to be nearby, they often appear at higher Galactic latitudes. Without significant dust extinction, we expect to detect CNe even at the largest Galactic distances, but we do not expect to detect even the brightest DNe outbursts beyond $\sim$ 6 kpc. Since CNe are more luminous, we are still able to detect them down to latitudes near $b = 0^{\circ}$, although dust extinction can obscure CNe at the lowest latitudes.

\subsection{Outburst vs.\ Quiescent Brightness}
\begin{figure*}
\begin{center}
 \includegraphics[width=\textwidth]{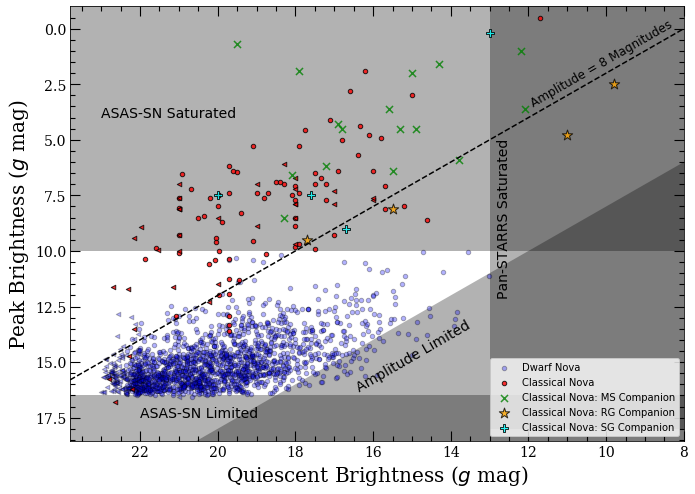}
\caption{Peak magnitude of outburst versus measured brightness in quiescence for all outbursts discussed in this work. Dwarf nova outbursts are shown in blue, and classical nova outbursts without companion information are shown in red. Those classical novae where the companion type is known are denoted as green X's, orange stars, and cyan crosses for main sequence, red giant, and sub-giant companions, respectively. The non-shaded region indicates where our analysis can measure outburst properties by combining ASAS-SN and Pan-STARRS observations. Only these surveys were used to study DNe, but AAVSO \textit{V}-band observations were utilized to study CNe that peak above the saturation limit of ASAS-SN. The dashed line shows an outburst amplitude of 8 mag. The ``amplitude-limited" diagonal shaded region shows the requirement that outbursts in our catalog must have amplitudes $>$2 mag, and quiescent brightness measurements in this region are likely contaminated by outbursts. The horizontal gray shaded region at top denotes the saturation limit of ASAS-SN (\textit{g} $\lesssim 10$ mag). The horizontal shaded region at bottom signifies two magnitudes brighter than the sensitivity limit of ASAS-SN (\textit{g} $\gtrsim 18$ mag). Finally, the vertical shaded region at right represents the saturation limit of the Pan-STARRS 3$\pi$ survey (\textit{g} $\lesssim 13$ mag). }
\label{Fig:peakqui}
\end{center}
\end{figure*}

Figure \ref{Fig:peakqui} shows the distribution of the sources by plotting peak outburst brightness against brightness in quiescence. The peak outburst magnitudes of CNe  are significantly brighter than for DNe, although this is largely due to selection effects. For DNe, the brightness in outburst were studied using data solely from ASAS-SN, so we do not include DN outbursts brighter than $\sim$10 mag (the saturation limit of ASAS-SN) in this study. The region with \textit{g} $\gtrsim$ 10\,mag, where ASAS-SN is saturated, is populated only with CNe because we rely on AAVSO observations to measure brighter peak magnitudes for CNe. 
We also note that ASAS-SN is sensitive to transients as faint as $\sim$18 magnitude, but since we are interested in measuring $t_2$, the outburst has to reach at least two magnitudes brighter than the survey magnitude limit. This detection range is shown in the non-shaded region in Figure \ref{Fig:peakqui}.
Objects that are not detected in quiescence are indicated by leftward facing triangles at the expected 98\% completeness limits. Where \cite{drb12} classified the companion of a CN, we have included the classification as main sequence (MS), red giant (RG), and sub-giant (SG) stars. Luminous companions may contribute significantly to the quiescent flux we measure (indeed, novae with giant companions are found to have bright quiescent magnitudes), which will lower the estimated outburst amplitude.

\subsection{Outburst amplitude vs.\ t$_2$}\label{sec:a_t2}
\begin{figure*}
\begin{center}
 \includegraphics[width=\textwidth]{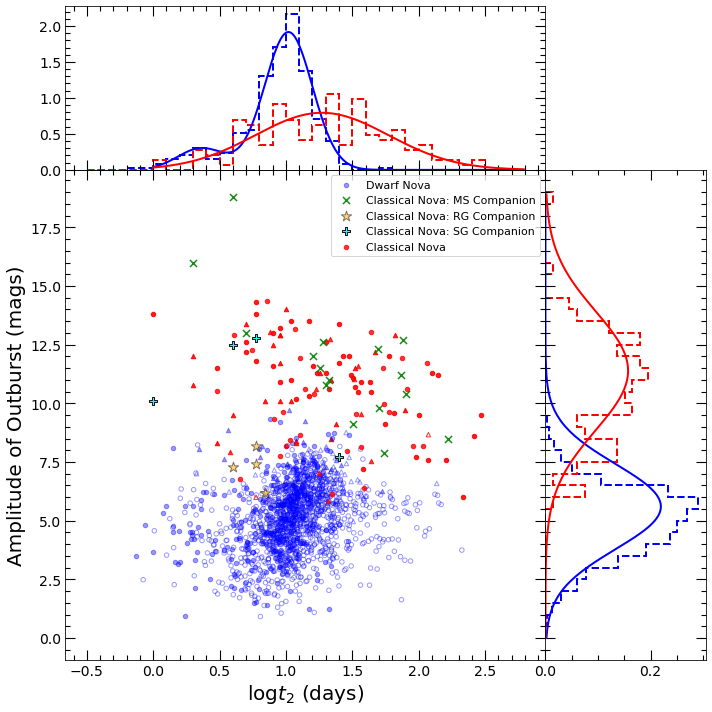}
 \caption{Amplitude of outburst versus the time t$_2$ to decline by two magnitudes from maximum for both CNe and DNe. 
A filled circle signifies that both the outburst amplitude and $t_2$ were estimated for that object. 
A triangle signifies that a lower limit was placed on the outburst amplitude, and 
an open symbol shows the upper limit that was placed on $t_2$. Though we are able to place lower and upper limits on  $t_2$, we only show the upper limit for visualization purposes. 
Blue objects denote DN outbursts, while CNe analyzed in this work and \protect\cite{ssh10} are represented with symbols as in Figure \ref{Fig:peakqui}.
The top and right panels show the distributions of $t_2$ and outburst amplitude, respectively, with DNe shown in blue and CNe shown in red. 
The dashed histograms show the distributions of only measured values, not limits, and the solid lines shows the fits to the measured values including the limits.}
\label{Fig:at2}
\end{center}
\end{figure*}

The amplitude of outburst is shown as a function of $\log_{10} \left ( t_2 \right )$ in Figure \ref{Fig:at2} for both CN and DN outbursts. As expected, the majority of DNe have smaller outburst amplitudes than CNe, although there is significant overlap for amplitudes of 5--10\,mag. We find that the outburst amplitudes and decline times of both samples are well fit by normal distributions, with the exception of the decline times of DNe. These distributions were fit using censored statistics, as a fraction of our estimates for the amplitude of outburst and $t_2$ are limits and are shown along with histograms of measured values, not including limits, in Figure \ref{Fig:at2}. For CNe, the normal distribution of the outburst amplitude has a mean and standard deviation of $\mu = 11.43 \pm 0.25$\,mag and $\sigma = 2.57 \pm 0.20$\,mag, respectively. This is in comparison with the amplitudes of DNe, where 
$\mu = 5.13 \pm 0.04$\,mag and
$\sigma = 1.55 \pm 0.03$\,mag. There is a roughly 15$\%$ overlap in the outburst amplitude distributions of CNe and DNe, suggesting that this property alone is not sufficient to distinguish the two classes of objects.

The mean of the outburst amplitude distribution for DNe presented in this paper is larger than other measurements found using transient survey data alone \citep{ckk16,mup15}.
With typical CCD dynamic ranges of about 5 magnitudes, it is difficult to detect both the transient peak and the quiescent system in a single
survey unless the amplitude is less extreme \citep{dgd14}. 
By combining ASAS-SN and Pan-STARRS observations, we are able to measure and place lower limits on outburst amplitudes 
as high as 12 magnitudes.

We do not include error bars on Figure \ref{Fig:at2} for visualization purposes, but the lower and upper bounds on t$_2$ for each DNe and CNe can be found in Tables \ref{table:dn_results} and \ref{table:cn_results}, respectively. We have not estimated the error on the outburst amplitude and expect systematics to dominate. The error on the outburst amplitude should be relatively small in most instances, but for some small fraction of objects, the outburst amplitude may be significantly underestimated. The peak of the outburst can be missed if the object declines rapidly or occurs during a time of lower temporal cadence by ASAS-SN. In addition, for DNe that outburst frequently, there is a chance that the quiescent magnitude we measure from Pan-STARRS data is contaminated by outbursts. A more detailed discussion of possible errors, the sensitivity of our analysis, and possible selection effects is provided in Section \ref{sec:selection}. In considering the results presented in Tables \ref{table:dn_results} and \ref{table:cn_results}, we encourage the reader to take these caveats into consideration.

Fitting a log-normal distribution to the CN decline times, we find a mean and standard deviation of $\left \langle t_2 \right \rangle$ = 18.7 $\pm$ 1.9 days and 3.2 $\pm$ 0.2 days, respectively. The distribution of $t_2$ values for DNe is not well fit by a single log-normal distribution, but can be described as a homoscedastic double-log-normal distribution, with mean values equal to 2.4 $\pm$ 0.2 days for 12$\%$ of the sample and  10.5 $\pm$ 0.2 days for the remaining 88$\%$ of the sample. The common standard deviation is 1.52 $\pm$ 0.02 days. Visual inspection of ASAS-SN images of these ``fast'' outbursts confirm that the transients are real. 

The bimodality of the outburst durations in SU UMa dwarf novae is well documented, with normal outbursts lasting a few days and superoutbursts lasting roughly two weeks  \citep{warner_1995,osaki_96}. In our analysis, we only measure $t_2$ of the brightest outburst, so we expect our sample to be biased towards superoutbursts rather than normal outbursts.
One explanation for a short outburst is that the heating wave fails to move fully throughout the accretion disk of the CV, and the unheated colder region pulls material from hotter regions of the disk, shutting down the outburst \citep{Smak84}. In the case of superoutbursts, the heating wave reaches the outer edge of the disk, causing the disk to remain hot for a longer amount of time.

In addition to finding that the distributions are well fit by Gaussians, we also find strong evidence of a relationship between the amplitude of outburst and $t_2$ for DNe, and modest evidence of an inverse correlation for CNe. We use censored statistics to measure a linear correlation of the form
\begin{equation}
\label{eq:at2}
    \log_{10}\left ( t_2 \right ) = \beta \left( \rm{Amp} - \left < Amp \right > \right ) + \alpha
\end{equation}
where $\rm{Amp}$ is the outburst amplitude and $\left < \rm{Amp} \right >$ is the mean of only the measured outburst amplitudes ($\left < \rm{Amp} \right >$ = 10.57 for CNe and $\left < \rm{Amp} \right >$ = 4.91 for DNe). For DNe, we exclude the subset of fast DNe ($\log_{10} \left ( t_2 \right ) < 0.4$), and we find a fit with $\alpha$ = 0.980 $\pm$ 0.005 and $\beta$ = 0.061 $\pm$ 0.004. This fit has a modest intrinsic scatter of $\sigma = 0.178 \pm 0.004$, and
the correlation is highly significant, roughly 10$\sigma$. 

We are unable to find a previous study of our observed correlation between amplitude and $t_2$ in DNe. However, \cite{oop16} studied the correlation between outburst duration (the total time of the outburst) and the amplitude of outburst for DNe. For normal outbursts of SU UMa stars they found no significant correlation between outburst duration and outburst amplitude. However, for superoutbursts, they did find evidence for a correlation. We make no distinction between subtypes of DN outbursts and expect this sample to contain a higher fraction of superoutbursts since our measurement is for the brightest observed outburst of an object since 2013 and excludes the faster outbursts from the fit.

\begin{figure}
\begin{center}
 \includegraphics[width=0.45\textwidth]{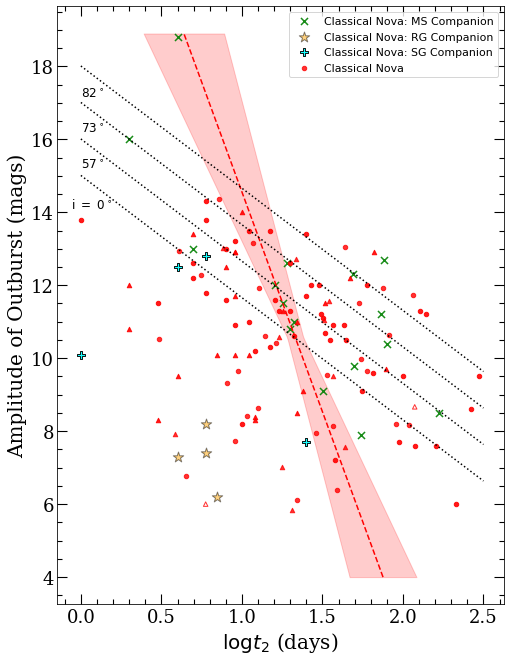}
 \caption{The outburst amplitude versus $\log{t_2}$ for the CNe analyzed in this work. The markers are the same as Figure \ref{Fig:at2}. The red dashed line and the red shaded region show our best-fit relation for CNe, with values and uncertainty given in \S \ref{sec:a_t2}. The dotted black lines show the expected theoretical correlation derived from the MMRD relationship from Figure 5.4 of \cite{warner_1995}.}
\label{Fig:at2_fit}
\end{center}
\end{figure}

For CNe, we find a less significant (roughly 3$\sigma$)  correlation with best-fit parameters to Equation \ref{eq:at2} of  $\alpha$ = 1.33 $\pm$ 0.05 and $\beta = -0.083 \pm 0.024$, and a large intrinsic scatter of $\sigma = 0.50\pm0.04$.
This fit is shown in Figure \ref{Fig:at2_fit} along with the predicted correlation derived from the MMRD relationship for various inclination angles  and assuming an absolute magnitude in quiescence of M$_V = 3.8$ \citep{warner_1995,cmd89}. Although the correlation for CNe is less significant than for DNe, the two populations have opposite slopes: amplitude and $t_2$ are anti-correlated for CNe, while they are positively correlated for DNe.

\cite{warner87} noted the substantial scatter in the relation between amplitude and $t_2$ for CNe. He attributed it to observational errors, a random distribution of inclination angles, and/or nova outbursts depending on multiple binary parameters (e.g., white dwarf mass, accretion rate, and core temperature), but the measured values appeared to be in general agreement with predictions.
\citet{Yaron_2005} point out that their models predict a population of low-amplitude CNe ($<$7 mag). Both \cite{warner87} and \citet{Yaron_2005} analyzed novae from the \citet{Duerbeck_1987} catalog where few low amplitude novae are present.
Our sample includes a larger population of such low-amplitude  CNe (many of which also have small $t_2$), likely due to newer surveys that are more sensitive to fainter and faster CNe. This serves to steepen the slope of the fit and further increase the variance around the anti-correlation between amplitude and $t_2$, compared to the results of \citet{warner87}. Our findings are consistent with the recent results from \cite{kck11} and \cite{sdl17}, who find a class of novae that deviate from the proposed MMRD relationship.

\section{Which Dwarf Novae might be mis-classified Classical Novae?}
\label{sec:hiding}
To test the idea that some CNe commonly get mis-characterized as DNe, we search for possible CN candidates in our sample of DN outbursts. 
This is likely the first large-sample analysis of this kind, but there is at least one example of an ASAS-SN transient that was initially characterized as 
a DN candidate but turned out to be a highly reddened CN 
(ASASSN-20ga; \citealt{dhk20}). Because of their high luminosities, Galactic CN eruptions can only appear faint (\textit{g} $\gtrsim$ 12\,mag) if they are affected by substantial dust extinction. For a Galactic transient to be a CN, it must have a peak absolute magnitude $M_{\rm g,peak}$ brighter than $-4.2$ mag ($M_{\rm g,peak} = -4.2$ mag is 3$\sigma$ fainter than the mean of the log-normal CN luminosity function presented in \citealt{shafter_2017}).
The absolute magnitude is tied to the peak apparent magnitude $m_{g,peak}$ by
\begin{equation}
    M_{\rm g,peak} = m_{\rm g,peak} - 5\log_{10} \left ( \frac{d}{10~\rm{pc}}\right ) - A_{g} ,
\label{Eq:AbsMag}
\end{equation}
where $d$ is the distance of the object in pc and $A_{g}$ is the amount of extinction. 
To place an upper limit on how luminous a given transient can possibly be, we take the peak apparent magnitude of the transient measured from ASAS-SN, an upper limit on the distance given reasonable constraints,
and the maximum \textit{g}-band extinction in the direction of the transient from \cite{sfd11}.

In our previous analysis, we only considered DNe in the Pan-STARRS field of view ($\delta > -30^{\circ}$), but here we apply those lessons learned to inspect all DN outbursts detected in ASAS-SN. This results in 2688 DNe with outbursts detected by ASAS-SN, spread over the entire sky.
For 1039 of the objects, parallaxes were measured with \emph{Gaia} at $\geq 3 \sigma$ significance, and for those, we use the 1$\sigma$ upper limits on the distances given in \cite{brf18}.
For those objects without significant distance estimates in \cite{brf18}, we used a Galactic upper limit of $d =$ 30 kpc. This is likely too conservative for any direction in the Galaxy, but a directional upper limit on the distance is beyond the scope of this work. At this time, we are more focused on being complete than robust when identifying candidates and plan to investigate a more reasonable Galactic distance upper limit as a function of position in Kawash et al.\ (2021, in preparation).

\begin{figure}
\begin{center}
 \includegraphics[width=0.48\textwidth]{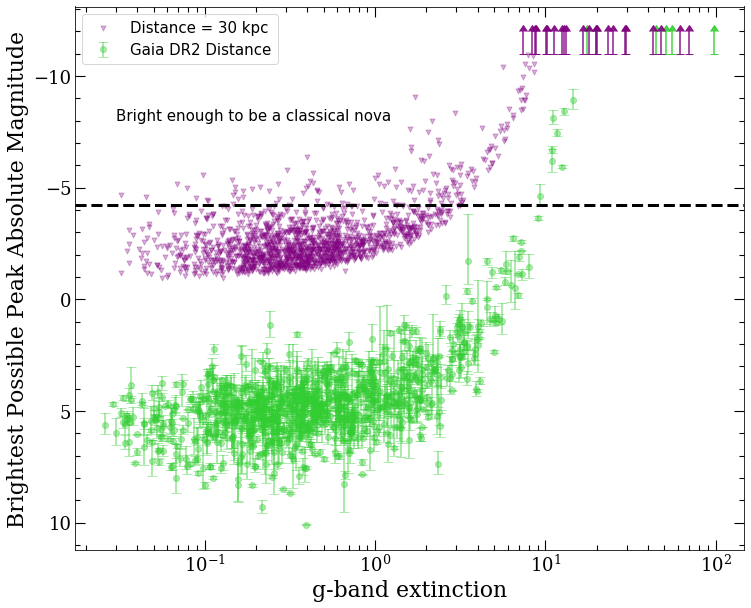}
 \caption{The brightest possible peak \textit{g}-band absolute magnitude an outburst could have while still being in the Galaxy, as a function of the amount of \textit{g}-band extinction. 
 The objects with 3$\sigma$ parallax detections in \textit{Gaia} are assumed to be at the distances in \protect\cite{brf18} and are plotted in green. 
 The violet points are objects that do not have significant \textit{Gaia} parallaxes, so we place a limit on the brightest peak absolute magnitude by assuming a maximum Galactic distance of 30 kpc.
 In order to be luminous enough to be a CN, the absolute magnitude needs to be brighter than M$_g \approx$ $-$4.2\,mag, and this cutoff is shown as the dashed black line. The peak absolute magnitude presented here is not an accurate measurement especially for objects with no reliable distance estimates and those with large amounts of dust extinction; the plotted values are only used to select CN candidates.}
\label{Fig:absMagvsAg}
\end{center}
\end{figure}

We find that 201 (< 10\%) objects classified as DNe in our sample could have $M_{\rm g,peak} \lesssim -4.2$ mag, if they were behind all of the dust estimated by \cite{sfd11}, as shown in Figure \ref{Fig:absMagvsAg}. To be clear, these are not exact peak absolute magnitude measurements, especially for objects with no reliable distances (shown in purple) and high extinction, and are only being used to identify CN candidates. We further rule out objects where we have measured the outburst amplitude to be $<$ 5\,mag, the lower bound of CN amplitudes based on Figure \ref{Fig:at2}, and objects with more than one detected outburst in an observing season (bounded by Solar conjunction). Though objects with multiple outbursts in ASAS-SN data are much more likely to be dwarf novae than recurrent novae, we can not rule out the latter. There are no known recurrent novae in the Galaxy that recur on timescales less than a decade, but M31 recurrent nova M31N 2008-12a erupts every year \citep{dh19}. If objects like this, dubbed `rapid recurrent novae,' exist in the Galaxy, they should be less luminous and evolve more quickly than a typical classical nova, making them easily confused with DNe. Therefore, we only eliminate objects with multiple outbursts in a year so our search is sensitive to rapid recurrent novae. Overall, we find that 94 objects have outburst amplitudes, recurrence times, and possibly luminosities consistent with that of a Galactic classical or recurrent nova.  

Many of these objects have been confirmed as DNe through an identification spectrum or a measurement of the superhump period, but we find that 27 sources were not confirmed through any method. Quiescent multi-band photometry of these remaining candidates 
can provide insights to the distance and ultimately constrain the luminosity of the transient. In the Galactic plane, a candidate that is relatively blue is likely close by, and therefore will have a lower luminosity at peak brightness, suggesting a DN outburst. Conversely, a candidate that is brighter in redder filters is likely reddened by dust, implying a higher luminosity and a CN outburst. This strategy---identifying highly reddened quiescent counterparts---is only possible for candidates within a few degrees of the Galactic plane and that can be securely matched to multi-band optical catalogs. As described in \S \ref{sec:amp}, we find quiescent counterparts in Pan-STARRS, and also add in coverage of southerly declinations by using the DeCAPS catalog \citep{sgl18} (cross matching for both catalogs is as explained in Section \ref{sec:catmatch}).
We use the $griz$ photometry to estimate reddening by fitting the observed spectral energy distributions of the CVs in question to de-reddened SDSS CV colors from \cite{kato12} by varying the amount of reddening according to extinction laws \citep{cjc89,mathis90}. For each of our candidates, this results in a distribution of extinction values that are plugged into the three-dimensional all-sky extinction map stitched together in \cite{bovy15} to find a range of plausible distances. This strategy only worked for 19 of the candidates: those that were able to be cross matched and in fields close to the plane, where the large amount of dust can constrain the distance. We find that all 19 are consistent with the peak luminosity of a dwarf nova, and zero are consistent with the peak luminosity of a classical nova. This analysis will also be useful to shed light on the nature of CV candidates that are discovered in the future, so it is made publicly available as an iPython notebook at \url{https://github.com/amkawash/CV_colors_luminosity}

\begin{figure*}
\begin{center}
 \includegraphics[width=0.9\textwidth]{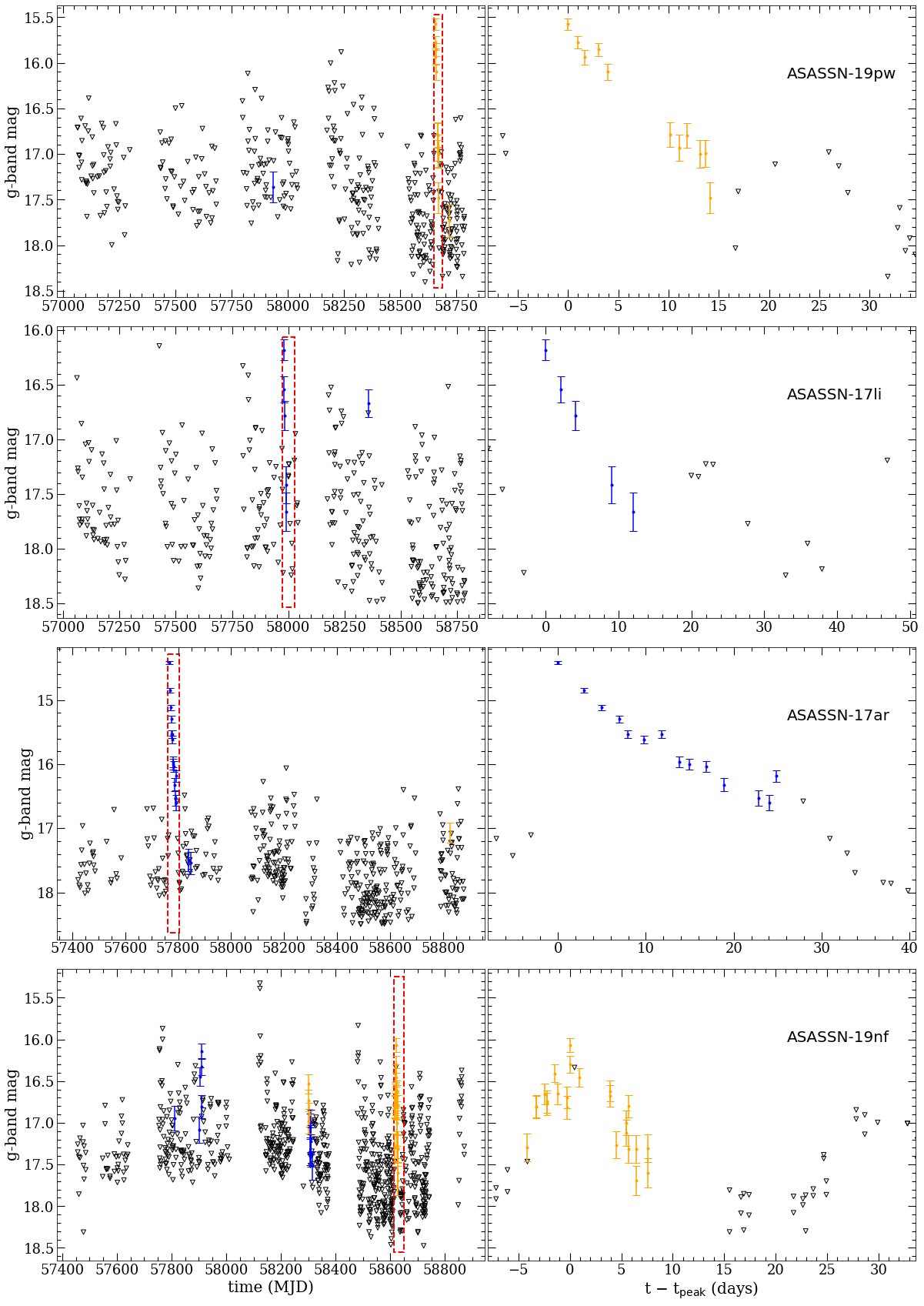}
 \caption{ASAS-SN Light curves for 4 of the candidates listed in Table \ref{table:cands}. The left column shows all observations of these objects and the right column shows the observations around the brightest outburst. Blue and orange points denote the $\geq$5-sigma detections from \textit{V}-band and \textit{g}-band observations, respectively, and the black triangles signify the $\geq$5-sigma upper limits from non-detections.}
\label{Fig:nova_cand_lc}
\end{center}
\end{figure*}

\begin{figure*}
\begin{center}
 \includegraphics[width=0.9\textwidth]{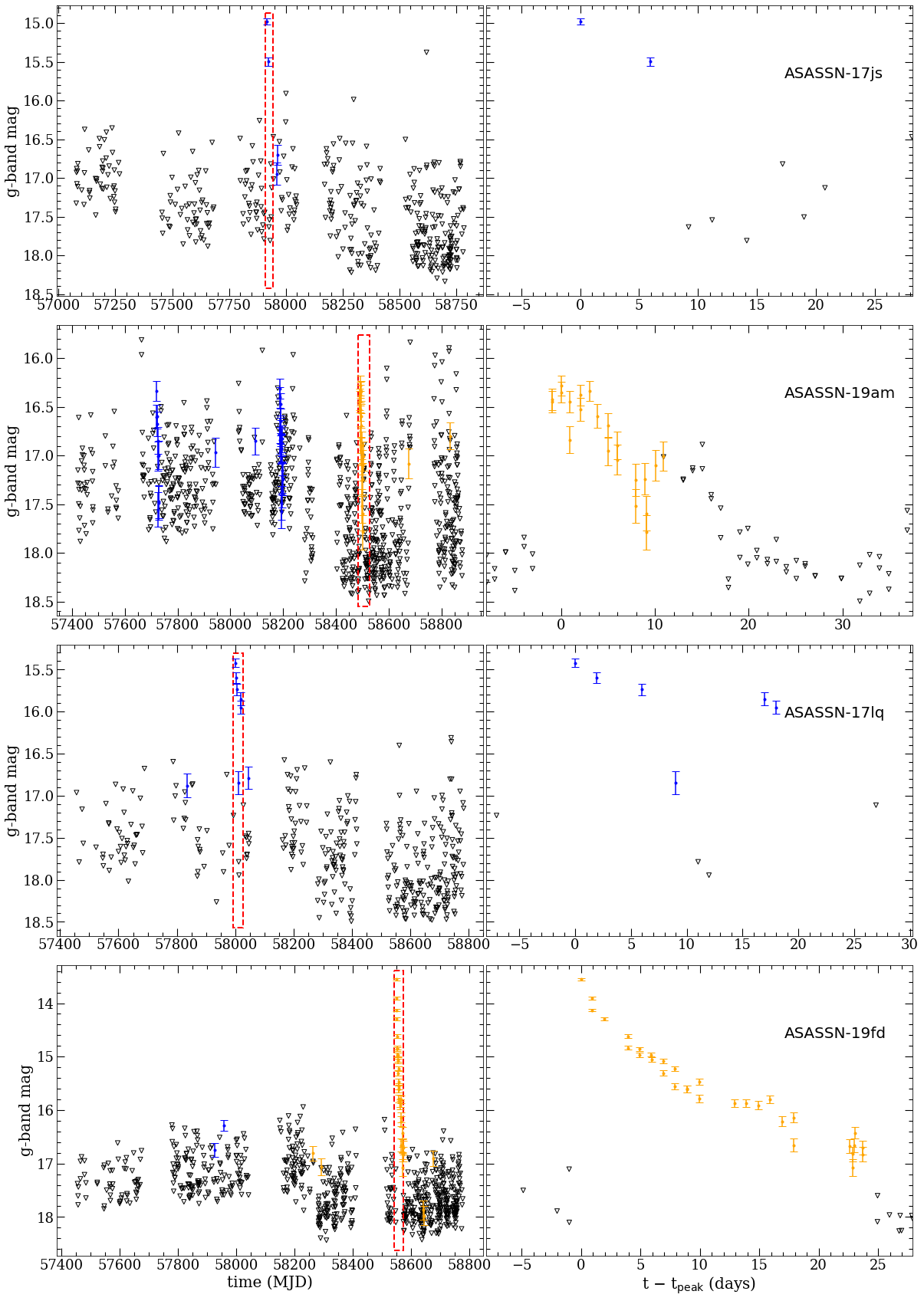}
 \caption{Same as Figure \ref{Fig:nova_cand_lc} for the remaining 4 candidates in Table \ref{table:cands}.}
\label{Fig:nova_cand_lc2}
\end{center}
\end{figure*}

So, for all but 8 of the 2688 DN outbursts detected by ASAS-SN, we find evidence suggestive of their DN nature. An identification spectrum is needed to determine if any of the 8 remaining candidates are mis-classified classical or recurrent novae. These candidates are shown in Table \ref{table:cands} listing equatorial and Galactic coordinates, peak \textit{g} apparent magnitude observed in the ASAS-SN light curve, a limit on outburst amplitude if able to be measured, t$_2$, extinction along the line of sight from \citet{sfd11}, an estimate of the outburst recurrence time measured from the ASAS-SN light curve ($\tau_R$), and if the outburst is luminous enough to be a CN as close as 10 kpc. The ASAS-SN light curves of all of these candidates are shown in Figures \ref{Fig:nova_cand_lc} and \ref{Fig:nova_cand_lc2}.

\begin{table*}
\centering
\caption{Classical and Rapid Recurrent Nova Candidates}
\begin{tabular}{lccccccccccc}
\hline
Name & Right Ascension & Declination & l & b & Peak  & Amp. & t$_2$ & A$_{g}$ & $\tau_R$ & 10 kpc\\ 
& (h~m~s)& ($^\circ$~$'$~$"$) & (degrees) & (degrees) & (mag) & (mag) & (days) & (mag) & (years) & (bool)\\
\hline
ASASSN-17li	& 18:38:22.00 & $-$09:43:47.4 & 22.790 & $-$1.551 & 16.2 & nan &[12.0 - 32.9] & 10.2  & >3 & Y\\
ASASSN-17ar & 10:01:11.14 &  $-$55:11:56.3 & 280.224 & $-$0.018 & 14.4 & nan & 20.7 & 7.6 &  >3 & Y\\
ASASSN-19nf & 14:19:35.09 & $-$59:58:24.0 & 313.755 & 1.033 & 16.1 & $>7.5$ & [7.6 - 15.5] & 19.8 & 0.9 & Y\\
ASASSN-19am & 09:30:39.31 & $-$54:47:04.3 & 276.609 & $-$2.521 & 16.3 & nan & [10.9 - 17.8] & 7.4 & 0.8 & Y\\
ASASSN-17lq & 17:29:28.81 & $-$38:02:26.8 & 350.512 & $-$2.042 & 15.4 & $>8.2$ & [9.0 - 11.0] & 10.1 & >3 & Y\\
ASASSN-19pw & 18:31:05.75 & $-$14:47:52.6 & 17.469 & $-$2.303 & 15.6 & 6.7 & [14.2 - 16.6] & 6.4 & >4 & Y\\
ASASSN-17js & 18:21;09.05 & $-$19:24:47.6 & 12.275 &  $-$2.347 & 15.0 & 6.5  & [5.9 - 9.2] & 6.9 & >3 & Y\\
ASASSN-19fd & 17:03:19.29 &  $-$29:52:23.3 & 354.045 & 7.099 & 13.5 & nan & 7.9 & 1.4 & >4 & N\\

\hline
\multicolumn{2}{l}{}
\end{tabular}
\label{table:cands}
\end{table*}


\section{Conclusions}
In this work, we characterize the brightest outburst of 1618 DN outbursts detected by ASAS-SN and  93 CNe observed by
ASAS-SN and AAVSO contributors. In general agreement with previous results, we find that the mean outburst amplitude of CNe is 11.4 magnitudes, significantly larger than the mean DN outburst of 5.1 magnitudes. However, we find significant overlap in their distributions, at the $\sim 15\%$ level. Although the outburst amplitude is a fairly good indicator for determining the nature of a CV outburst, it is clear that a CV outburst with amplitude in the range 5$-$10 mag is ambiguous in nature. Similarly, the mean decline time, or $t_2$, of CNe is larger than that of DNe, but a majority of the distributions overlap, especially at lower values. Because there is an overlap in parameter space between DN outbursts and CN eruptions, we have presented a technique to identify CN versus DN candidates based solely on photometric data. This will be a necessary tool to handle the large number of CV outbursts discovered in the LSST era.

The primary motivation for this work was driven by the possibility that CNe are being mis-characterized as DNe. To explore this prospect, we have investigated every DN that declines by two magnitudes from maximum in ASAS-SN. The majority of outbursts are inconsistent with the luminosity of a CN, but there is a small fraction that could be bright enough if they are behind most of the dust along the line of sight. We looked into this subset and found that only 8 objects (out of 2688) are still ambiguous based on available data. A classification spectrum will be needed to confirm if any of these candidates are CNe characterized as DNe, but it is clear that there is no significant number of Galactic classical novae hiding in the large sample of dwarf novae. The transient community appears to be doing an effective job classifying CV outbursts.

Our results suggest that either Galactic nova rate predictions are too high or there must be other factors than classical nova mis-classification causing the discrepancy between reported and predicted classical novae. Recent observations from  Palomar Gattini-IR have revealed a sample of highly reddened and optically missed novae due to Galactic extinction \citep{dkh21}. We plan to explore to what degree interstellar dust has an effect on ASASN's, and other optical observer's, ability to discover classical novae.

\section*{Acknowledgments}

This research has made use of the International Variable Star Index (VSX) database, operated at AAVSO, Cambridge, Massachusetts, USA. We acknowledge with thanks the variable star observations from the \emph{AAVSO International Database} contributed by observers worldwide and used in this research.

A.K., L.C., E.A., and K.V.S.\ acknowledge financial support of NSF award AST-1751874 and a Cottrell fellowship of the Research Corporation. J.S.\ acknowledges support from the Packard Foundation.
BJS, CSK, and KZS are supported by NSF grant AST-1907570. CSK and KZS are supported by NSF grant AST-181440. 

We thank the Las Cumbres Observatory and its staff for its continuing support of the ASAS-SN project. ASAS-SN is supported by the Gordon and Betty Moore Foundation through grant GBMF5490 to the Ohio State University, and NSF grants AST-1515927 and AST-1908570. Development of ASAS-SN has been supported by NSF grant AST-0908816, the Mt. Cuba Astronomical Foundation, the Center for Cosmology and AstroParticle Physics at the Ohio State University, the Chinese Academy of Sciences South America Center for Astronomy (CAS- SACA), and the Villum Foundation. 

The analysis for this work was performed primarily in \texttt{ipython} \citep{pg07} using-  \texttt{numpy} \citep{oliphant2006guide,van2011numpy}, \texttt{Astropy} \citep{astropy:2018}, \texttt{Matplotlib}  \citep{Hunter:2007}, and \texttt{scipy} \citep{vgo20}. The PanSTARRS1 Catalog was accessed using packages from \texttt{MAST CasJobs}\footnote[3]{\url{http://casjobs.sdss.org/CasJobs}} developed by the JHU/SDSS team.


\bibliographystyle{aasjournal}
\bibliography{biblio}



\appendix
\renewcommand\thefigure{\thesection.\arabic{figure}}    
\setcounter{figure}{0}    

\renewcommand\thetable{\thesection.\arabic{table}}    
\setcounter{table}{0}    
\section{Appendix}

\subsection{Filter Transformations}
\label{sec:filter}
The photometry utilized in this work makes use of a range of blue-green filters: 
the \textit{V} filter used by AAVSO observers, the ASAS-SN \textit{V} filter, the ASAS-SN \textit{g} filter, the Pan-STARRS \textit{g}$_{P1}$ filter, and the Gaia \textit{G}$_{BP}$ filter. Although all of these filters are centered around a similar wavelength, the flux of a source in each filter band can be different, especially for reddened objects. To account for this, all \textit{V}-band observations were transformed to \textit{g}-band using
\begin{equation}
\label{eq:vgr}
    \textit{V} - \textit{g} = -0.017 -0.508 \times \left ( \textit{g} - \textit{r} \right ),
\end{equation} 
when \textit{g}$_{P1}$ and \textit{r}$_{P1}$ observations were available \citep{kb18}. 
Pan-STARRS only provides estimates of colors in quiescence, so to transform \textit{V}-band data
in outburst to \textit{g}-band, we assume
typical CV colors in quiescence (\textit{B}$-$\textit{V} = 0.1; \citealt{bruch84}) and typical color changes during outburst ($\Delta$(\textit{B}$-$\textit{V}) = $-$0.1 for DNe and $\Delta$(\textit{B}$-$\textit{V}) = 0.13 for CNe; \citealt{warner_1995}). These rough color estimates were converted from \textit{B} and \textit{V} to \textit{g} and \textit{r} using equation \ref{eq:vgr} and additional filter transformations from \cite{kb18}. This implies an intrinsic color of \textit{g}$-$\textit{r} = $-$0.29 for DNe and $-$0.08 for CNe. We then use the measured Pan-STARRS colors to estimate the reddening and ultimately the observed \textit{g} $-$ \textit{r} color in outburst.

For objects without color information in Pan-STARRS, we have to make additional assumptions. For DNe, we assume \textit{g}$_{P1}-$\textit{r}$_{P1}$ = 0 in quiescence, but for CNe we assume a \textit{g}$_{P1}-$\textit{r}$_{P1}$ = 1, as most CNe are more distant and closer to the Galactic plane, and therefore reddened by dust. Although all of these color estimates and assumptions are very crude, this transformation only changes the flux of a typical object by a fraction of a magnitude. However, for some CNe with high reddening, the transformation applied can exceed a magnitude. If the reddening is substantially underestimated, the error in the magnitude shift could be as high as a magnitude, making the source appear brighter than it actually was in quiescence.

\textit{G}$_{BP}$ observations of CNe in quiescence were corrected as
\begin{equation}
        \textit{g}-\textit{G}_{BP} = -0.318 + 0.932 x -0.932 x^2 
    +0.507 x^3 -0.107 x^4 +0.007 x^5
\end{equation}
where x = \textit{G}$_{BP} - \textit{G}_{RP}$ using a polynomial we fit to sources with colors ranging from \textit{G}$_{BP}$ $-$ \textit{G}$_{BP}$ = 0.6 to \textit{G}$_{BP}$ $-$ \textit{G}$_{BP}$ = 3.4 in both \textit{Gaia} and SDSS. This correction is typically a fraction of a magnitude, reaching up to $\sim$1 magnitude for the most highly reddened objects. We assume any differences between the ASAS-SN \textit{g} filter, the Pan-STARRS \textit{g} filter, and the SDSS \textit{g} filter are negligible. 

\subsection{Quiescent Magnitude Measurements} \label{sec:quimag}

The Pan-STARRS $3\pi$ Steradian Survey \citep{cmm16} covers the sky north of declination $\delta = -30^{\circ}$, and the stacked catalog has a median 5$\sigma$ depth of \textit{g}$_{P1}\simeq 23.3$ mag. Quiescent magnitude measurements were made using the \textit{g}$_{P1}$ passband, as this filter is the most similar to the \textit{g}-band measurements made by ASAS-SN and avoids the need for any extinction correction to the outburst amplitude.

The Pan-STARRS 3$\pi$ catalog reaches its full depth and astrometric accuracy by combining 12 exposures taken between 2009 Jun 2 and 2014 Mar 31 \citep{cmm16}. This is a potential issue for estimating the quiescent magnitude of DNe, because a substantial fraction of them have multiple outbursts over this time frame. Therefore, it is possible, and in many cases highly probable, that the flux of a DNe listed in the stack or mean catalog is contaminated by outbursts and does not accurately reflect the quiescent brightness of the source. To minimize the possibility of this issue, flux measurements were obtained from the \texttt{ForcedWarpMeasurement} table from Pan-STARRS Data Release 2. This table contains single epoch forced photometry measurements at the position of objects detected in the stacked images \citep{fmc16}. For each source, the quiescent magnitude was estimated from the median flux from the faintest 50$\%$ of \textit{g}$_{P1}$ observations. This increases the chance that observations contaminated by outbursts were excluded, and we only measure the quiescent brightness for objects where the median absolute deviation of the apparent magnitude measurements is less than 0.9 mag to avoid estimates that are still likely contaminated.

If no source is present in the Pan-STARRS catalog (details of cross matching in Section~\ref{sec:catmatch}), an upper limit was placed on the brightness based on the \textit{g}$_{P1}$ magnitude corresponding to the 98$\%$ completeness of the field obtained from the $\texttt{StackDetEffMeta}$ table. The limiting magnitudes are estimated from the number of fake sources recovered for each skycell in the 3$\pi$ stacked survey \citep{msd11}.

We estimate the quiescent magnitude of the CNe in the same way for those in the observing field of Pan-STARRS. Since the CNe in our sample only outburst once, we are able to supplement the Pan-STARRS photometry with \emph{Gaia} DR2 photometry for CNe that erupted after \emph{Gaia} DR2 observations \cite{bvp18}, for objects outside of the observing field of Pan-STARRS. Quiescent magnitudes for CNe estimated using \emph{Gaia} were made using the $G_{BP}$ filter, as it is the most similar to ASAS-SN \textit{g}-band, and transformed to \textit{g} as describe in \S\ref{sec:filter}. 

\subsection{Astrometry and Catalog Matching}
\label{sec:catmatch}
When matching sources from different catalogs, the possibility exists that two different sources with similar sky positions can be mistaken as the same object; this largely depends on the astrometric accuracy of the surveys involved. \cite{ygo19, ygo19b} investigated the astrometric accuracy of various surveys, including ASAS-SN, by comparing the positions of objects reported by an individual survey to those independently reported by 
the \textit{Gaia} Alerts Project\footnote[4]{\url{http://gsaweb.ast.cam.ac.uk/alerts/home}}. They estimated the positional accuracy of various surveys and found that 95$\%$ of discoveries by ASAS-SN have astrometric errors $<$ 3.4 arcseconds. Since many of the objects in our sample are discovered by ASAS-SN, we use a positional offset threshold of four arcseconds when considering matches between the DNe in our sample and objects in Pan-STARRS. 
This is a generous search radius, as the typical error on the discovery position is  $\sim$1 arcsecond \citep{jks18}. For CNe, we assume the discoveries are followed up at higher angular resolution, and therefore place a stricter bound on the astrometric uncertainty of one arcsecond. 

We use CN and DN coordinates as listed in the VSX catalog. When transients discovered by ASAS-SN and other surveys are entered into VSX, their positions are updated using \emph{Gaia} DR2 (taking epoch and equinox J2000.0) if there is an object within a fraction of an arcsecond from the reported transient position. If no object exists within $\sim$1 arcsecond, other optical surveys with similar limits like Pan-STARRS, SDSS \citep{sloan00}, and the Guide Star Catalog (GSC2.3; \citealt{llb08}) are checked for matches within a fraction of an arcsecond. If there are no matches in surveys with reliable astrometry, the position is derived from the discovery report or follow-up astrometry. 
 
For each DN in our sample, we estimate the probability that it is coincident by chance with a different, nearby object in the Pan-STARRS catalog. 
A positional offset was defined for each source as the angle between the DN's VSX position and the closest object in the Pan-STARRS stack catalog. 
We then search the Pan-STARRS catalog at random positions on a circle of one degree around the source and compute the frequency of having a Pan-STARRS source closer than the measured positional offset. 
Only sources that had random matches less than 5\% of the time were considered secure matches. The maximum positional offset that results in a secure match is roughly 2 arcseconds.

For CNe, we only consider a Pan-STARRS object within 1 arcsecond of the VSX position to be a secure match.
DNe with a Pan-STARRS source within 4 arcseconds and CNe with a source within 2 arcseconds where random matches were found $>$5\% of the time are considered ambiguous matches, and we do not attempt to estimate their outburst amplitudes. 
If no source exists in the Pan-STARRS catalog within 4 arcseconds for DNe and 2 arcseconds for CNe of the VSX position, we consider the quiescent counterpart definitively not detected in the Pan-STARRS 3$\pi$ stack catalog, 
and place an upper limit on the quiescent brightness.  We successfully estimate or place a limit on the quiescent brightness for $\sim$90\% of DNe in the Pan-STARRS observing field.

For CNe south of $\delta = -30$ degrees, we consider \textit{Gaia} DR2 sources within one arcsecond to be secure matches. If no \emph{Gaia} DR2 source appears 
to be within two arcseconds of the CNe position, we place an upper limit on the brightness of the source. 
The magnitude that corresponds to 98$\%$ completeness in \textit{Gaia} is not currently available and likely spatially variable, 
but we assume this will roughly be the magnitude that corresponds to a magnitude error of 0.1\,mag. 
We find that this value is on average  $G_{BP}$ $\approx$ 19.7\,mag, and therefore use this as an upper limit on the magnitude for CNe with no \emph{Gaia} DR2 source within two arcseconds. By using Pan-STARRS and \textit{Gaia} observations we are able to measure or place a limit on $\sim$80\% of CN outbursts.

\subsection{Sensitivity/Contamination} \label{sec:selection}
\begin{figure*}
\begin{center}
 \includegraphics[width=0.98\textwidth]{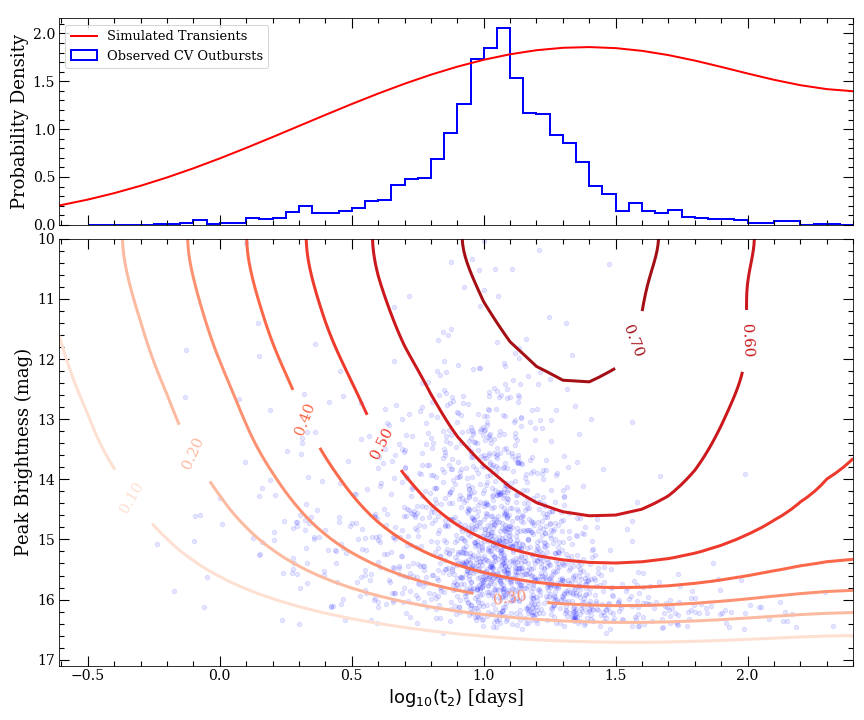}
\caption{Top: Distribution of observed time to decline by two magnitudes for detected DN outbursts in blue and the probability distribution for measuring the decline as a function of $t_2$ in red. Both distributions are normalized so that the area under the curve is unity, but the red curve is then multiplied by a factor of four for visualization purposes.
Bottom: Peak \textit{g} magnitudes of DN outbursts vs.\ measurements and limits of the time to decline by two magnitudes from maximum. Real DN outbursts are shown as blue circles. The red contours show the fraction of time the decline time of fake transients could be successfully measured or constrained with an upper limit in our analysis.}
\label{Fig:sim_at2}
\end{center}
\end{figure*}

We injected fake transients into our data and attempted to recover them in order to estimate the ranges of peak magnitudes and decline times to which our analysis is sensitive, given the cadence of ASAS-SN. We generated linearly declining (in magnitude versus time) outbursts with peak apparent magnitudes ranging from 18 to 10\,mag. and $t_2$ values ranging from an hour to a year. These outbursts were injected into the ASAS-SN light curves obtained for our CV sample with random outburst epochs. We sampled the mock-outburst evolution with the cadence and sensitivity of observed ASAS-SN light curves.
The same analysis that was run on the real CV sample was run on the fake transients, in order to estimate how frequently we could successfully measure or place limits on $t_2$. The results are shown in Figure \ref{Fig:sim_at2} and are compared to the measured values from our CV sample.

The top panel in Figure \ref{Fig:sim_at2} shows the distributions of real CV decline times as a blue histogram, and the relative frequency with which fake transient decline times could be estimated (red line). This shows that our analysis is best at detecting transients with $t_2 \approx$ 30 days, although we most frequently find DN outbursts that decline by two magnitudes from maximum in roughly 10 days. Also, the sensitivity of our analysis as a function of decline time drops off much more slowly than the distribution of real decline times. Though we are certainly worse at detecting CV outbursts with extremely short decline times, the results are not significantly biased by our analysis and observing cadence.

The bottom panel of Figure \ref{Fig:sim_at2} shows the distribution of peak magnitude versus $t_2$ for the real DN outbursts (shown in blue) and contours of the probability that the fake transients were recovered (shown in red). Even the brightest and slowest transients are not always recovered successfully, and this is largely due to an outburst happening while Sun constrained. The completeness is not strongly dependent on the peak magnitude of the source unless it is near the limiting magnitude of ASAS-SN.

As discussed in the main text, there is a chance that the quiescent magnitude measured from Pan-STARRS is contaminated by an outburst. To investigate the likelihood for this to occur, we measure the outburst duty cycle, defined as the fraction of time a CV spends in outburst. We estimate this as the number of days the object is detected by ASAS-SN divided by the total number of days the field is observed. Since image subtraction light curves were used to study the DN outbursts in our sample, the assumption that each detection is during an outburst is a safe one, but it can break down when contamination from a nearby bright star occurs. Due to this, we only estimate the duty cycle for objects with no \textit{g} $<$ 14 mag stars within an 8 ASAS-SN pixel (64 arcseconds) radius.

\begin{figure}
\begin{center}
 \includegraphics[width=0.48\textwidth]{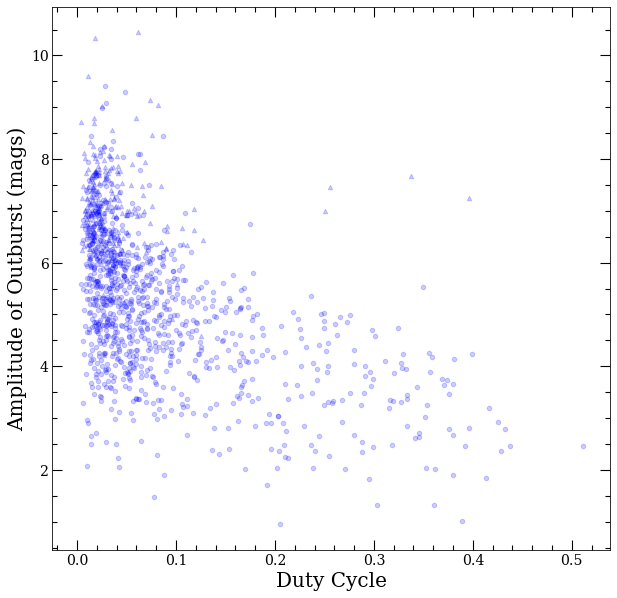}
\caption{Outburst amplitude for the dwarf novae as a function of duty cycle.}
\label{Fig:duty}
\end{center}
\end{figure}

As shown in Figure \ref{Fig:duty}, it does appear to be the case that DNe with larger duty cycles have smaller outburst amplitudes. Though this is consistent with the nature of DNe \citep{ckk16}, our analysis may significantly underestimate the outburst amplitude for an object with a high duty cycle. Objects that spend more time in outburst have a higher probability of being observed by Pan-STARRS in outburst, and thus will result in an underestimate in the outburst amplitude estimated in this work. This should not significantly alter the distribution of the outburst amplitude, but we encourage the use of caution when quoting the outburst amplitude of an individual object, 

Contamination in the ASAS-SN light curve from a nearby, bright star could result in a false positive of a DN outburst. Careful inspection and flagging of the light curves was performed to mitigate these artifacts, though the possibility still remains. 

\subsection{Classical and Reccurent Nova Outburst Properties}
\label{sec:nova_results}

\begin{figure*}
\begin{center}
 \includegraphics[width=1.0\textwidth]{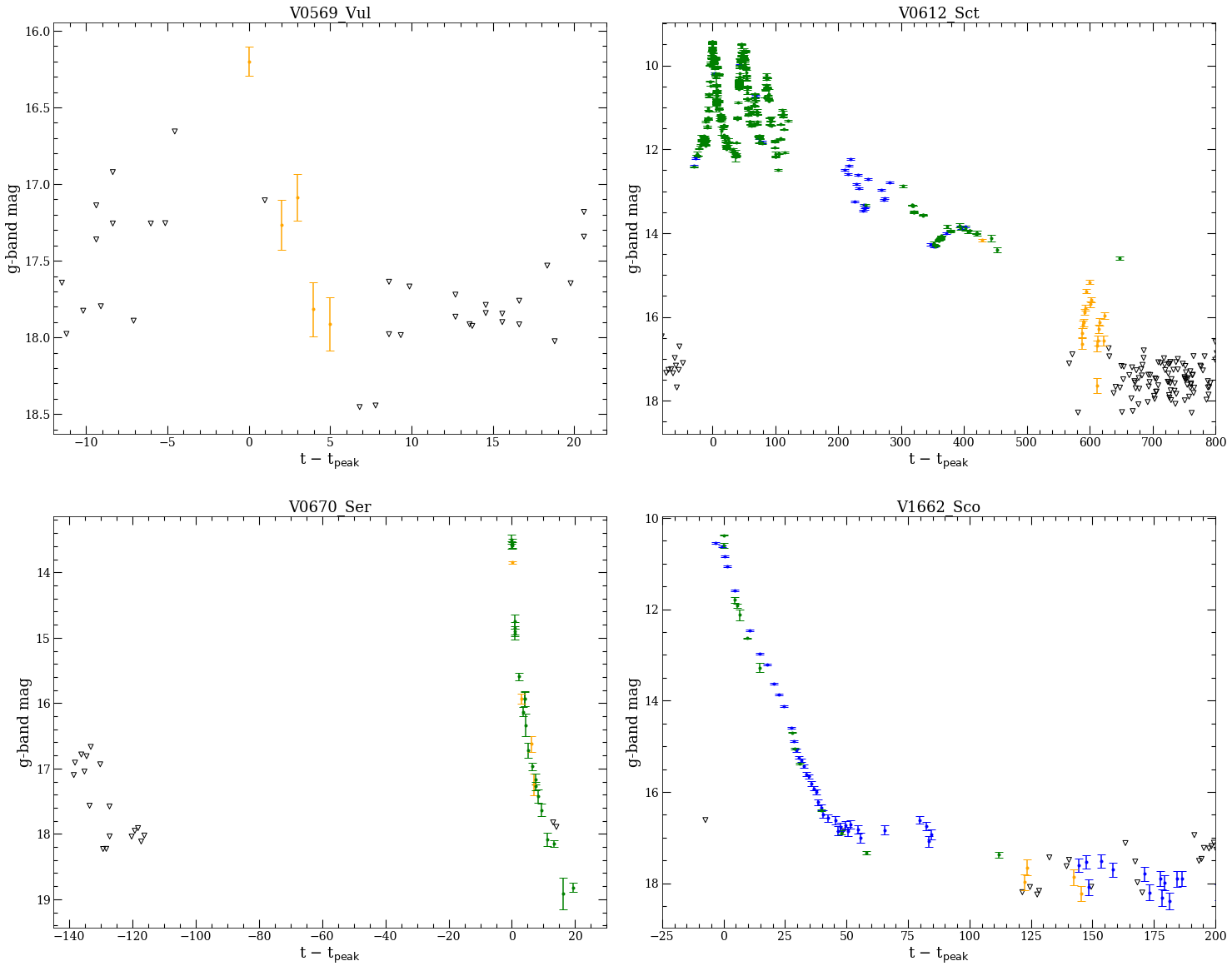}
\caption{Light curves of four classical novae analyzed in this work. The $\geq$5-sigma detections for ASAS-SN \textit{g}-band, ASAS-SN \textit{V}-band and AAVSO \textit{V}-band observations are shown in orange, blue, and green respectively after converting to brightness in \textit{g}-band. The black triangles denote 5-sigma upper limits from non-detections These show examples of faint outbursts (top left), flares causing multiple peaks (top right), outbursts during solar conjunction (bottom left), and smooth declines (bottom right).}
\label{Fig:nova_lcs}
\end{center}
\end{figure*}

Here, we provide the outburst properties of the classical novae and recurrent novae in our sample in Table \ref{table:cn_results}. This table is available electronically in a machine readable format. The data used to measure these properties for four classical novae are shown as light curves in Figure \ref{Fig:nova_lcs}. 

\begin{deluxetable*}{lccccccccc}
\tabletypesize{\small}
\tablecolumns{10}
\tablewidth{0pt}
\tablecaption{Outburst Properties of Classical and Recurrent Novae \label{table:cn_results}}
\tablehead{
\colhead{Name} & 
\colhead{RAJ2000} & 
\colhead{DEJ2000} & 
\colhead{Peak} &
\colhead{Amp.} & 
\colhead{Amp. Flag} & 
\colhead{t$_2$} & 
\colhead{t$_2$ Flag} &
\colhead{t$_{2,low}$} & 
\colhead{t$_{2,up}$}\\
\colhead{ } & \colhead{h~m~s} & 
\colhead{$^\circ$~$'$~$"$} & 
\colhead{mag} &
\colhead{mag} & 
\colhead{boolean} & 
\colhead{days} & 
\colhead{boolean} &
\colhead{days} & 
\colhead{days} }
\startdata
V0392 Per & 4:43:21.37 & 47:21:25.9 & 7.0 & 10.5 & 1 & 3.0 & 1 & 3.0 & 3.1 \\
V0339 Del & 20:23:30.68 & 20:46:03.8 & 4.6 & 13.2 & 1 & 11.8 & 1 & 11.8 & 12.5 \\
V2659 Cyg & 20:21:42.32 & 31:03:29.4 & 9.9 & 11.7 & 1 & 115.5 & 1 & 114.7 & 116.4 \\
V0569 Vul & 19:52:08.25 & 27:42:20.9 & 16.2 & 6.0 & 0 & 6.0 & 0 & 5.0 & 6.8 \\
V0962 Cep & 20:54:23.75 & 60:17:06.9 & 11.6 & 11.1 & 0 & 32.5 & 1 & 31.2 & 34.1 \\
V0435 CMa & 7:13:45.84 & $-$21:12:31.3 & 10.4 & 11.5 & 1 & 53.4 & 1 & 49.3 & 55.3 \\
V2860 Ori & 6:09:57.45 & 12:12:25.2 & 10.6 & 9.6 & 1 & 9.4 & 1 & 9.0 & 10.0 \\
V5668 Sgr & 18:36:56.83 & $-$28:55:40.0 & 4.4 & 11.9 & 1 & 75.3 & 1 & 74.5 & 77.8 \\
V5855 Sgr & 18:10:28.29 & $-$27:29:59.3 & 8.4 & 11.9 & 1 & 12.7 & 1 & 7.3 & 16.3 \\
V5856 Sgr & 18:20:52.25 & $-$28:22:12.1 & 6.5 & 14.4 & 1 & 7.2 & 1 & nan & 14.5 \\
V1707 Sco & 17:37:09.54 & $-$35:10:23.2 & 12.9 & 6.8 & 1 & 4.5 & 1 & 3.6 & 5.6 \\
V1659 Sco & 17:42:57.68 & $-$33:25:42.9 & 13.6 & 6.1 & 1 & 22.0 & 1 & 20.7 & 22.6 \\
V3661 Oph & 17:35:50.41 & $-$29:34:23.8 & 12.3 & 7.9 & 0 & 3.8 & 1 & 2.1 & 5.6 \\
V5669 Sgr & 18:03:32.77 & $-$28:16:05.3 & 9.5 & 9.5 & 1 & 33.6 & 1 & 33.3 & 41.9 \\
V5853 Sgr & 18:01:07.78 & $-$26:31:43.4 & 12.9 & 8.2 & 1 & 36.7 & 1 & 36.6 & 37.6 \\
V5667 Sgr & 18:14:25.15 & $-$25:54:34.7 & 10.0 & 10.0 & 1 & 54.5 & 1 & 54.1 & 55.1 \\
V3662 Oph & 17:39:46.10 & $-$24:57:55.8 & 14.7 & 7.6 & 0 & 43.7 & 1 & 43.0 & 47.3 \\
V3890 Sgr & 18:30:43.29 & $-$24:01:08.9 & 8.1 & 12.9 & 1 & 4.1 & 1 & 4.1 & 4.1 \\
V5666 Sgr & 18:25:08.76 & $-$22:36:02.6 & 10.1 & 8.6 & 1 & 12.7 & 1 & 12.2 & 13.0 \\
V0612 Sct & 18:31:45.86 & $-$14:18:55.5 & 9.4 & 10.6 & 1 & 13.9 & 1 & 13.9 & 13.9 \\
V0613 Sct & 18:29:22.93 & $-$14:30:44.2 & 11.6 & 9.5 & 0 & 36.8 & 1 & 36.2 & 37.1 \\
V3665 Oph & 17:14:02.53 & $-$28:49:23.3 & 10.0 & 11.6 & 0 & 34.8 & 1 & 33.5 & 35.0 \\
V3666 Oph & 17:42:24.11 & $-$20:53:08.6 & 9.4 & 12.7 & 0 & 21.6 & 1 & 21.5 & 23.2 \\
V2944 Oph & 17:29:13.42 & $-$18:46:13.8 & 9.6 & 10.4 & 1 & 16.2 & 1 & 16.0 & 16.7 \\
V5857 Sgr & 18:04:09.45 & $-$18:03:55.8 & 11.7 & 10.6 & 0 & 16.9 & 1 & 15.7 & 17.4 \\
V0670 Ser & 18:10:42.29 & $-$15:34:18.0 & 13.5 & 8.7 & 0 & 118.3$^a$ & 0 & 1.0 & 118.3 \\
V0659 Sct & 18:39:59.70 & $-$10:25:41.9 & 8.9 & 13.0 & 0 & 7.6 & 1 & 7.3 & 8.3 \\
V1830 Aql & 19:02:33.38 & 3:15:19.0 & 16.8 & 5.8 & 0 & 20.6 & 1 & 20.6 & 20.6 \\
V1831 Aql & 19:21:50.15 & 15:09:24.8 & 15.8 & 7.0 & 0 & 17.8 & 1 & 17.5 & 18.7 \\
V0906 Car & 10:36:15.42 & $-$59:35:53.7 & 6.5 & 13.1 & 1 & 43.7 & 1 & 43.0 & 44.5 \\
V0549 Vel & 8:50:29.62 & $-$47:45:28.3 & 9.7 & 8.2 & 1 & 90.1 & 1 & 90.0 & 92.7 \\
FM Cir & 13:53:27.59 & $-$67:25:00.9 & 6.7 & 10.6 & 1 & 82.3 & 1 & 81.2 & 82.9 \\
V1405 Cen & 13:20:55.35 & $-$63:42:19.1 & 11.3 & 8.2 & 1 & 108.4 & 1 & 102.5 & 108.8 \\
V1655 Sco & 17:38:19.31 & $-$37:25:08.7 & 12.0 & 8.0 & 1 & 28.9 & 1 & 28.8 & 28.9 \\
V1662 Sco & 16:48:49.62 & $-$44:57:03.2 & 10.4 & 9.3 & 1 & 8.1 & 1 & 6.6 & 9.6 \\
V1657 Sco & 16:52:18.87 & $-$37:54:18.9 & 13.3 & 6.4 & 1 & 38.7 & 1 & 36.0 & 40.9 \\
V1656 Sco & 17:22:51.46 & $-$31:58:37.1 & 12.0 & 7.7 & 1 & 9.0 & 1 & 8.9 & 9.5 \\
V1661 Sco & 17:18:06.37 & $-$32:04:27.7 & 11.3 & 8.4 & 1 & 10.7 & 1 & 10.0 & 13.6 \\
V0408 Lup & 15:38:43.86 & $-$47:44:42.1 & 10.4 & 9.7 & 1 & 59.4 & 1 & 58.1 & 60.7 \\
V0407 Lup & 15:29:01.79 & $-$44:49:39.5 & 7.4 & 12.3 & 1 & 5.5 & 1 & 5.1 & 5.6 \\
\enddata
\tablenotetext{a}{The eruption likely occurred during solar constraint and t$_{2,up}$ is the time from before solar constraint to once the light curve dropped below the apparent two magnitude threshold.}
\tablecomments{Names, positions, peak apparent brightness, amplitude of outburst, and time it takes to decline to decline two magnitudes from maximum brightness of classical and recurrent novae in our sample. The Amp. Flag column equals one when we are able to make a measurement of the outburst amplitude and zero when we are able to place a lower limit. The t$_2$ Flag column equals one when were are able to detect the object below the two magnitude threshold and is zero when there is only a non-detection below this threshold or when the eruption appeared to occur during solar constraint. The t$_{2,low}$ column gives the time until last detection brighter than the two magnitude threshold and the t$_{2,up}$ column gives the time until the first detection or non-detection fainter this threshold. These last two columns are lower and upper limits on t$_2$, respectively. }
\end{deluxetable*}




\label{lastpage}
\end{document}